\documentstyle[12pt,epsfig]{article}

\newskip\humongous \humongous=0pt plus 1000pt minus 1000pt

\newif\ifdtup

\def\pr#1{#1^\prime}

\def\beq{\begin{equation}}
\def\eeq{\end{equation}}

\def\beqn{\begin{eqnarray}}
\def\eeqn{\end{eqnarray}}
\relax

\def\dotx{\dotx{\dot\overline{x}}}

\relax

\catcode`\@=11

\@addtoreset{equation}{section}
\def\theequation{\thesection\arabic{equation}}

\def\@normalsize{\@setsize\normalsize{15pt}\xiipt\@xiipt
\abovedisplayskip 14pt plus3pt minus3pt%
\belowdisplayskip \abovedisplayskip
\abovedisplayshortskip \z@ plus3pt%
\belowdisplayshortskip 7pt plus3.5pt minus0pt}

\def\small{\@setsize\small{13.6pt}\xipt\@xipt
\abovedisplayskip 13pt plus3pt minus3pt%
\belowdisplayskip \abovedisplayskip
\abovedisplayshortskip \z@ plus3pt%
\belowdisplayshortskip 7pt plus3.5pt minus0pt
\def\@listi{\parsep 4.5pt plus 2pt minus 1pt
     \itemsep \parsep
     \topsep 9pt plus 3pt minus 3pt}}

\@twosidetrue
\relax

\catcode`@=12

\catcode`\@=11

\def\section{\@startsection{section}{1}{\z@}{3.5ex plus 1ex minus
   .2ex}{2.3ex plus .2ex}{\large\bf}}

\def\thesection{\arabic{section}.}

\def\appendix{\setcounter{section}{0}
 \def\thesection{APPENDIX \Alph{section}:}
 \def\theequation{\Alph{section}.\arabic{equation}}}
\def\ps@headings{\def\@oddfoot{}\def\@evenfoot{}
\def\@oddhead{\hbox{}\hfill
 \makebox[.5\textwidth]{\raggedright\ignorespaces --\thepage{}--
 \hfill {}}}  %instead of {\rm FERMILAB--Pub--\FERMIPUB}}}
\def\@evenhead{\@oddhead}
\def\subsectionmark##1{\markboth{##1}{}}
}

\ps@headings

\catcode`\@=12

\def\figcap{\section*{Figure Captions\markboth
 {FIGURECAPTIONS}{FIGURECAPTIONS}}\list
 {Fig. \arabic{enumi}:\hfill}{\settowidth\labelwidth{Fig. 999:}
 \leftmargin\labelwidth
 \advance\leftmargin\labelsep\usecounter{enumi}}}
 \relax
\def\tablecap{\section*{Table Captions\markboth
 {TABLECAPTIONS}{TABLECAPTIONS}}\list
 {Table \arabic{enumi}:\hfill}{\settowidth\labelwidth{Table 999:}
 \leftmargin\labelwidth
 \advance\leftmargin\labelsep\usecounter{enumi}}}
 \relax
\def\reflist{\section*{References\markboth
 {REFLIST}{REFLIST}}\list
 {[\arabic{enumi}]\hfill}{\settowidth\labelwidth{[999]}
 \leftmargin\labelwidth
 \advance\leftmargin\labelsep\usecounter{enumi}}}
 \relax

\catcode`\@=11

\def\ps@headings{\def\@oddfoot{}\def\@evenfoot{}
\def\@oddhead{\hbox{}\hfill
 \makebox[.5\textwidth]{\raggedright\ignorespaces --\thepage{}--
 \hfill {}}}    %instead of {\rm FERMILAB--Pub--\FERMIPUB}}}
\def\@evenhead{\@oddhead}
\def\subsectionmark##1{\markboth{##1}{}}
}

\ps@headings

\catcode`\@=12

\relax

\catcode`\@=11
\def\prm{\fam \z@}
\catcode`\@=12

\relax
\def\pl#1#2#3{{\it Phys. Lett. }{\bf #1}(19#2)#3}
\def\zp#1#2#3{{\it Z. Phys. }{\bf #1}(19#2)#3}
\def\prl#1#2#3{{\it Phys. Rev. Lett. }{\bf #1}(19#2)#3}

\def\pr#1#2#3{{\it Phys. Rev. }{\bf #1}(19#2)#3}
\def\np#1#2#3{{\it Nucl. Phys. }{\bf #1}(19#2)#3}

\def\app#1#2#3{{\it Acta Phys. Polon. }{\bf #1}(19#2)#3}

\def    \hepph  #1 {{\tt hep-ph/#1}}
\def    \hepex  #1 {{\tt hep-ex/#1}}

\relax

  \newcommand{\ccaption}[2]{
    \begin{center}
    \parbox{0.85\textwidth}{
      \caption[#1]{\small{\it{#2}}}
      }
    \end{center}
    }
\begin{document}            
\newcommand\sss{\scriptscriptstyle}
\newcommand\me{m_e}
\newcommand\as{\alpha_{\sss S}}         
\newcommand\aem{\alpha_{\rm em}}
\newcommand\refq[1]{$^{[#1]}$}
\renewcommand\topfraction{1}       % Max. Fraz. di pagina per float in t
\renewcommand\bottomfraction{1}    % Max. Fraz. di pagina per float in b
\renewcommand\textfraction{0}      % Min. Fraz. di pagina per testo 
\setcounter{topnumber}{5}          % Max # float in position t
\setcounter{bottomnumber}{5}       % Max # float in position b
\setcounter{totalnumber}{5}        % Max # float in same page
\setcounter{dbltopnumber}{2}       % Max # large float
\newsavebox\tmpfig
\newcommand\settmpfig[1]{\sbox{\tmpfig}{\mbox{\ref{#1}}}}
%
%\/\/\/\/\/\/\/\/\/\/\/\/\/\/\/\/\/\/\/\/\/\/\/\/\/\/\/\/\/\/\/\/\/\/\/\/\
\begin{titlepage}
\nopagebreak
\vspace*{-1in}
{\leftskip 11cm
\normalsize
\noindent   
\newline
GEF-TH-7/1997 \newline
hep-ph/9707478

}
\vskip 1.0cm
\vfill
\begin{center}
{\large \sc Target mass effects}
\\
{\large \sc in polarized deep-inelastic scattering}
\vfill
\vskip .6cm 
{\bf Andrea Piccione}
\vskip .2cm
{Dipartimento di Fisica, Universit\`a di Genova, Italy}
\vskip .5cm                                               
{\bf Giovanni Ridolfi}
\vskip .2cm
{INFN Sezione di Genova, Genoa, Italy}

\end{center}
\vfill
\nopagebreak
\begin{abstract}
{\small
We present a computation of nucleon mass corrections to nucleon structure
functions for polarized deep-inelastic scattering. We perform a fit to
existing data including mass corrections at first order in $m^2/Q^2$
and we study the effect of these corrections on physically interesting
quantities. We conclude that mass corrections are generally small,
and compatible with current estimates of higher twist
uncertainties, when available.
}
\end{abstract}
\vskip0.7truecm
\end{titlepage}

\section{Introduction}
Experimental information on deep-inelastic scattering of polarized leptons off
different kinds of polarized nucleon targets has become more and more accurate
in the past few years~\cite{EMC}-\cite{E154}. This accumulated knowledge,
combined with recent theoretical progress in the computation
of perturbative QCD
quantities relevant to polarized deep-inelastic scattering \cite{NLO}, has
allowed a next-to-leading order determination of the polarized parton
densities,
using data on the structure function $g_1(x,Q^2)$ which determines the cross
section asymmetries in the case of longitudinally polarized leptons
and nucleons in the
Bjorken limit~\cite{GehrmannStirling}-\cite{ABFR}.
Reasonably good determinations of
the strong coupling constant $\as$ and of the axial vector coupling $g_A$ have
also been performed~\cite{ABFR}. 

A large part of experimental data in polarized deep-inelastic scattering are
taken at relatively low values of $Q^2$. In particular, $Q^2$ is usually around
1~GeV$^2$ for data points in the small-$x$ region, which is particularly
interesting because there the effects of $Q^2$ evolution are more important
(data at even lower values of $Q^2$ are also available, but they are usually not
included in perturbative analyses). In this kinematical region, contributions
suppressed by inverse powers of $Q^2$ can become important. These contributions
can be of two different origins. There are power-suppressed terms arising from
the operator product expansion of the hadronic tensor $W^{\mu\nu}$. These terms
originate from matrix elements of operators of non-leading twist (they are
usually referred to as dynamical higher twists). Their effect is not controlled
by perturbation theory, and it is very difficult to assess their importance. A
second class of power-suppressed contributions originates from taking into
account the finite value of the nucleon mass $m$ in the kinematics of the
leading-twist cross section. These corrections can be computed exactly, and have
been studied in detail in the past~\cite{GP} in the case of unpolarized
deep-inelastic scattering. Knowledge of kinematic effects is of course necessary
in order to extract information on dynamical higher twists from experimental
data. 

The problem of calculating kinematic higher twist terms in polarized
deep-inelastic scattering has been considered in ref.~\cite{MU}.\footnote{The
results of ref.~\cite{MU} have been used in ref.~\cite{KU} to compute target
mass corrections to the Bjorken sum rule.} There, the reduced matrix elements
$a_n,d_n$ of the relevant operators in the OPE were expressed in terms of
polarized structure functions, taking mass corrections into account; these
expressions reduce to moments of the structure functions in the massless limit,
but do not have a simple parton model interpretation in the case $m\neq 0$. For
this reason, the result of ref.~\cite{MU} is not directly applicable in a full
analysis of polarized deep-inelastic scattering data. 

In this paper, we calculate target mass corrections in polarized deep-inelastic
scattering extending the analogous work of ref.~\cite{GP} for the unpolarized
case. This has the advantage of yielding the final result in a form which is
appropriate for phenomenological applications, that is, we will express moments
of polarized structure functions as functions of the reduced operator matrix
elements. The details of our calculations are presented in Sect.~2. An important
point is of course the interplay between dynamical and kinematical higher
twists; in Sect.~2 we discuss this point in some detail. In Sect. 3, we use
mass-corrected formulae in an analysis of existing data in the framework of QCD
at next-to-leading order, and compare the results with those obtained in the
massless limit; in particular, we will compare our results with those of
ref.~\cite{ABFR}, where an estimate is given of the uncertainties coming from
higher twist effects on the determination of physically interesting quantities.
We will see that the effect of mass corrections is indeed within the
uncertainties estimated in ref.~\cite{ABFR}. Finally, we present our conclusions
in Sect.~4. 

\section{Calculation}
Our calculation follows closely the analogous one in the case of unpolarized
deep-inelastic scattering, performed by Georgi and Politzer back in
1976~\cite{GP}. We will adopt the usual notation for the definition of
kinematical quantities in deep-inelastic scattering, used for example in
ref.~\cite{kodaira}. In the formalism of the operator product expansion, the
antisymmetric part $T_{\mu\nu}^A$ of the forward amplitude relevant for
deep-inelastic scattering, 
\beq
T_{\mu\nu}(p,q,s)=
i\int d^4x e^{iqx} <p,s\mid T\left[J_\mu(x)J_\nu(0)\right]\mid p,s>
\eeq
is given by~\cite{kodaira}
\beqn
T_{\mu\nu}^A&=&-i\sum_{n=1}^{\infty}
\frac{[1-(-1)^n]}{2}
\left(\frac{2}{Q^2}\right)^n q_{\mu_1} \ldots q_{\mu_{n-2}}
\nonumber \\
&&\sum_{i}\Bigg[\epsilon_{\mu\nu\lambda\sigma}q^{\lambda}q_{\mu_{n-1}}
E_{1, i}^n(Q^2,\as)<p,s\mid R_{1, i}^{\sigma\mu_1 \ldots \mu_{n-1}}\mid p,s>
\nonumber\\
&&+ Q_{\mu\nu\lambda\sigma}\frac{n-1}{n}E_{2, i}^n(Q^2,\as)
<p,s\mid R_{2, i}^{\lambda\sigma\mu_1 \ldots \mu_{n-2}}\mid p,s>
\Bigg],
\label{tmunu}
\eeqn
where $J_\mu$ is the electromagnetic current, and
$Q_{\mu\nu\lambda\sigma}$ is defined as
\beq
Q_{\mu\nu\lambda\sigma}=
\epsilon_{\mu\rho\lambda\sigma}q_\nu q^{\rho}-\epsilon_{\nu
\rho\lambda\sigma}q_\mu q^{\rho}-q^2\epsilon_{\mu\nu\lambda\sigma}.
\eeq
The operators $R_{1,i}^{\sigma\mu_1\ldots\mu_{n-1}}$ are given by
\beqn
&&R_{1,i}^{\sigma\mu_1 \ldots \mu_{n-1}}=i^{n-1}\left[\bar{\psi}\gamma_{5}
    \gamma^{\sigma}D^{\mu_1} \ldots D^{\mu_{n-1}}\frac{\lambda_{i}}{2}
    \psi \right]_S   \;\;\;\; i=1,\ldots,8;
\label{R1NS}
\\
&&R_{1,\psi}^{\sigma\mu_1...\mu_{n-1}} =
i^{n-1}\left[\bar{\psi}\gamma_{5}\gamma^{\sigma}D^{\mu_1}\ldots
D^{\mu_{n-1}}\psi \right]_S;
\label{R1S}
\\
&&R_{1,G}^{\sigma\mu_1...\mu_{n-1}} =
i^{n-1}Tr~\left[\epsilon^{\sigma\alpha\beta\gamma}F_{\beta\gamma}
D^{\mu_1}...D^{\mu_{n-2}}F_{\alpha}^{\mu_{n-1}}\right]_S
\label{R1G}.
\eeqn
Here $D^\mu$ is the QCD covariant derivative, and the symbol $[\ldots]_ S$
means complete symmetrization in the indices $\sigma,\mu_1,\ldots,\mu_{n-1}$;
$F_{\mu\nu}$ is the usual QCD gluon tensor.
The twist-3 operators $R_{2,i}^{\lambda\sigma\mu_1 \ldots \mu_{n-2}}$
are given by
\beqn
&&R_{2,i}^{\lambda\sigma\mu_1...\mu_{n-2}} =
i^{n-1} \left[\bar{\psi}\gamma_{5}\gamma^{\lambda}D^{\sigma}
D^{\mu_1}...D^{\mu_{n-2}}\frac{\lambda_{i}}{2}\psi \right]_{S'},
\;\;\;\; i=1,\ldots,8; 
\label{R2NS}
\\
&&R_{2,\psi}^{\lambda\sigma\mu_1...\mu_{n-2}} =
i^{n-1} \left[ \bar{\psi}\gamma_{5}\gamma^{\lambda}D^{\sigma}D^{\mu_1}...
D^{\mu_{n-2}}\psi \right]_{S'}
\label{R2S}
\\
&&R_{2, G}^{\lambda\sigma\mu_1...\mu_{n-1}} =
i^{n-1}Tr ~\left[ \epsilon^{\sigma\alpha\beta\gamma}F_{\beta\gamma}
D^{\mu_1}...D^{\mu_{n-2}}F_{\alpha}^{\lambda} \right]_{S'},
\label{R2G}
\eeqn
where $[\ldots ]_{S'}$ indicates antisymmetrization with respect to
$\lambda$ and $\sigma$ and symmetrization with respect to other indices.
The coefficient functions $E_{1,i}^n(Q^2,\as)$ and $E_{2,i}^n(Q^2,\as)$
have been computed up to order $\as^2$ in perturbative 
QCD~\cite{kodairacf}.
The expansion of the forward scattering amplitude in powers of $m^2/Q^2$
is independent of the perturbative expansion in powers of $\as$; we can
therefore perform our calculation at leading order in $\as$, and then
insert perturbative corrections to coefficient functions in the final result. 
For $\as=0$ we have simply $E_{1,i}=E_{2,i}=1$ for $i=1,\ldots,8,\psi$
and $E_{1,G}=E_{2,G}=0$.

The matrix elements of the operators (\ref{R1NS}-\ref{R1S}) can be written as
\beq
<p,s\mid R_1^{\sigma\mu_1 \ldots \mu_{n-1}}\mid p,s>
 = -2ma_n M_1^{\sigma\mu_1 \ldots \mu_{n-1}}
\label{twisttwo}
\eeq
(we have omitted the index $i$, which is no longer necessary at leading order).
The tensor $M_1^{\sigma\mu_1 \ldots \mu_{n-1}}$ is the most general
rank-$n$ symmetric tensor which can be formed with one spin four-vector $s$ and 
$n-1$ momentum four-vectors $p$; furthermore, it must satisfy the 
tracelessness conditions
\beq
g_{\mu_i \mu_j} M_1^{\mu_1 \ldots \mu_n}=0
\eeq
for all pairs $i,j$. With these requirements, we find
\beq
M_1^{\mu_1 \ldots \mu_n}=\frac{1}{n}\sum_{j=0}^
        {\frac{n-1}{2}}\frac{(-1)^j}{2^j}\frac{(n-j)!}{n!}~
       \underbrace{g \ldots g}_j ~\left[sp \ldots p \right]_S~(m^2)^j
\label{M1}
\eeq
up to an overall normalization, which can be absorbed in the definition of
the reduced matrix elements $a_n$. The symbol
\beq
\underbrace{g \ldots g}_j
\eeq
represents a product of $j$ metric tensors $g^{\mu_l\mu_k}$, with
indices chosen among $\mu_1, \ldots, \mu_n$
in all possible ways; the remaining $n-2j$ indices of 
$M_1^{\mu_1 \ldots \mu_n}$ are carried by the symmetric product
$\left[sp \ldots p \right]_{S}$.
When the nucleon mass is neglected, only the first term of the sum,
$\left[s^{\sigma}p^{\mu_1}...p^{\mu_{n-1}}\right]_{S}$,
is retained; this is the 
standard result, used for example in ref.~\cite{kodaira}.

Consider now the twist-3 operators of eqs.~(\ref{R2NS}-\ref{R2G}). Their matrix
elements can be written as
\beq
<p,s\mid R_2^{\lambda\sigma\mu_1...\mu_{n-2}}\mid p,s> = 
md_n M_2^{\lambda\sigma\mu_1 \ldots \mu_{n-2}},
\label{twistthree}
\eeq
where the tensor $M_2$ must be antisymmetric in ($\lambda,\sigma$),
symmetric in all other indices, and traceless. It is easy to prove that
\beq
M_2^{\lambda\sigma\mu_1 \ldots \mu_{n-1}}=
\frac{n+1}{n}(s^\sigma\Pi^{\lambda\mu_1 \ldots \mu_{n-2}}-
              s^\lambda\Pi^{\sigma\mu_1 \ldots \mu_{n-2}}) 
+\frac{n-1}{n}(p^\sigma M_1^{\lambda\mu_1 \ldots \mu_{n-2}}-
               p^\lambda M_1^{\sigma\mu_1 \ldots \mu_{n-2}})
\eeq
where
\beq
\Pi^{\mu_1 \ldots \mu_n}=\sum_{j=0}^{\frac{n}{2}}
                   \frac{(-1)^j}{2^j}\frac{(n-j)!}{n!}~
                 \underbrace{g \ldots g}_j ~p \ldots p ~(m^2)^j
\label{Pi}
\eeq
is the most general rank-$n$ symmetric, traceless tensor that can be formed
with the momentum $p$ alone. For $m^2=0$, one recovers the usual result
\beq
<p,s\mid R_2^{\lambda\sigma\mu_1...\mu_{n-2}}\mid p,s>= 
md_n\left(s^\sigma p^\lambda-s^\lambda p^\sigma\right)
p^{\mu_1} \ldots p^{\mu_{n-2}}.
\eeq

The reduced matrix elements $a_n,d_n$ contain all the information on the proton
spin structure; they are related to moments of polarized structure functions.
Our next step consists in obtaining these relationships in the general case
$m\not=0$. To do this, we compute explicitly the amplitude $T_{\mu\nu}^A$ using
our results, eq.~(\ref{twisttwo}) and eq.~(\ref{twistthree}). We decompose
$T_{\mu\nu}^A$ into a twist-2 and a twist-3 component, 
\beq
T_{\mu\nu}^A=T^{(a)}_{\mu\nu}+T^{(d)}_{\mu\nu}.
\eeq
Let us first consider the contribution of twist-2 operators,
\beq
T_{\mu\nu}^{(a)}=2im\epsilon_{\mu\nu\lambda\sigma}q^\lambda
\sum_{n~odd}\left(\frac{2}{Q^2}\right)^n
q_{\mu_1} \ldots q_{\mu_{n-1}}
M_1^{\sigma \ldots \mu_{n-1}}a_n.
\eeq
Using eq.(\ref{M1}) and recalling that $x=Q^2/(2p \cdot q)$ we find
\beq
T_{\mu\nu}^{(a)}=\frac{2im}{p \cdot q}
\epsilon_{\mu\nu\lambda\sigma}q^{\lambda}
\sum_{n~odd}x^{-n}
\frac{a_n}{n^2}\sum_{j=0}^{\frac{n-1}{2}}
\left(\frac{x^2 m^2}{Q^2}\right)^j
\frac{(n-j)!}{j!(n-1-2j)!}
\left(s^{\sigma}+(n-2j-1)\frac{s \cdot q}{p \cdot q}p^\sigma\right).
\eeq
Following ref.~\cite{GP}, we change summation index from $n$ to $l$, with
$n=2l+2j+1$, and exchange the summation order of $l$ and $j$. This gives
\beq
T_{\mu\nu}^{(a)} = \frac{2im}{p \cdot q}
\epsilon_{\mu\nu\lambda\sigma}q^{\lambda}
     \sum_{l=0}^{\infty}x^{-(2l+1)}
     \left(s^\sigma+2l\frac{s \cdot q}{p \cdot q} p^\sigma\right)
     \sum_{j=0}^{\infty}\frac{a_{2l+2j+1}}{(2l+2j+1)^2}
     \left(\frac{m^2}{Q^2}\right)^j
\frac{(2l+j+1)!}{j!(2l)!}.
\label{ta}
\eeq
We now define functions $F_{a,d}(x)$ by
\beq
a_n = \int_0^1 dy ~y^n ~F_a(y); \;\;\;\;\ d_n = \int_0^1 dy ~y^n ~F_d(y).
\label{fdef}
\eeq
It is easy to prove that
\beqn
&&\frac{a_n}{n} = \int_0^1 dy ~y^{n-1} ~G_a(y); \;\;\;\;
  \frac{a_n}{n^2} = \int_0^1 dy ~y^{n-1} ~H_a(y) 
\label{gha} \\
&&\frac{d_n}{n} = \int_0^1 dy ~y^{n-1}~G_d(y);  \;\;\;\;
  \frac{d_n}{n^2} = \int_0^1 dy ~y^{n-1}~H_d(y),
\label{ghd}
\eeqn
where
\beq
G_{a,d}(x)=\int_{x}^1 dy F_{a,d}(y); \;\;\;\;
H_{a,d}(x)=\int_{x}^1 \frac{dy}{y} G_{a,d}(y).
\eeq
These definitions allow us to perform the summation over $j$ in eq.~(\ref{ta});
in fact, we can write
\beqn
T_{\mu\nu}^{(a)} &=& \frac{2im}{p \cdot q}
\epsilon_{\mu\nu\lambda\sigma}q^{\lambda}
     \sum_{l=0}^{\infty}x^{-(2l+1)}
(2l+1)\left(s^\sigma+2l\frac{s \cdot q}{p \cdot q}p^\sigma\right)
\nonumber \\
&&    \int_0^1 dy ~y^{2l} ~H_a(y)  
    \sum_{j=0}^{\infty}\left(\frac{y^2 m^2}{Q^2}\right)^j 
\frac{(2l+j+1)!}{j!(2l+1)!}.
\eeqn
The sum over $j$ can now be performed using the identity
\beq
\frac{1}{(1-z)^{n+1}} = \frac{1}{n!}\frac{d^n}{dz^n} \frac{1}{1-z}= 
                      \frac{1}{n!} \sum_{j=0}^{\infty} \frac{(n+j)!}
                          {j!} z^j,
\label{serie}
\eeq
with the result
\beq
T_{\mu\nu}^{(a)} = \frac{2im}{p \cdot q}
\epsilon_{\mu\nu\lambda\sigma}q^{\lambda}
    \sum_{n~odd} x^{-n}
    n\left(s^\sigma-(1-n)\frac{s \cdot q}{p \cdot q}p^\sigma\right) 
\int_0^1 dy ~\frac{y^{n-1}}{\left(1-\frac{y^2 m^2}{Q^2}\right)^{n+1}}~H_a(y),
\label{afinal}
\eeq
where we have defined $n=2l+1$.
The same procedure applied to the twist-3 term yields
\beqn
T_{\mu\nu}^{(d)}&=&\frac{2im}{p \cdot q}
\epsilon_{\mu\nu\lambda\sigma}q^\lambda
      \sum_{n=3, 5, \ldots}^{\infty}
     x^{-n}
      \int_0^1 dy ~\frac{y^{n-1}}
                        {\left(1-\frac{y^2m^2}{Q^2}\right)^n} 
\nonumber \\
&&\left\{
  (n-1)\left[G_d(y)~s^\sigma-nH_d(y)\frac{s\cdot q}{p\cdot q}p^\sigma\right]
  + 2n\frac{y^2 m^2}{Q^2} 
      \frac{G_d(y)+ H_d(y)}{1-\frac{y^2m^2}{Q^2}}s^\sigma
      \right\}
\nonumber \\
\label{dfinal}
\eeqn

It is now possible to obtain the $n^{th}$ moments of the polarized
structure functions $g_1$ and $g_2$.
The antisymmetric part of the hadronic tensor $W_{\mu\nu}$ is defined by
\beq
iW^A_{\mu\nu}=\frac{1}{\pi}{\rm Im}T^A_{\mu\nu}.
\eeq
From the analytic structure of $T^A_{\mu\nu}$
in the complex $\nu\equiv p\cdot q$ plane, one can prove that
\beq
T^A_{\mu\nu}=
\frac{2}{\pi}\sum_{n=1,3,\ldots}^{\infty} x^{-n}\int_0^1dy y^{n-1}
{\rm Im}T^A_{\mu\nu}=
\sum_{n=1,3,\ldots}^{\infty} x^{-n}\int_0^1 dy y^{n-1}~2iW^A_{\mu\nu}.
\eeq
In other words, the coefficient of $x^{-n}$ in $T^A_{\mu\nu}$
gives twice the $n^{th}$ moment of the hadronic tensor $iW^A_{\mu\nu}$.

On the other hand, $iW^A_{\mu\nu}$ is usually parametrized in the following
way:
\beq
iW^A_{\mu\nu}=
\frac{im}{p\cdot q}\epsilon_{\mu\nu\lambda\sigma}q^\lambda
\left[g_1(x,Q^2) s^\sigma
      +g_2(x,Q^2)\left(s^\sigma-\frac{q\cdot s}{p\cdot q}p^\sigma\right)
\right].
\eeq
We can therefore identify the $n^{th}$ moment of $g_1+g_2$ and $g_2$
as twice the coefficients of $s^\sigma$ and $-p^\sigma(s\cdot q)/(p\cdot q)$
in eqs.~(\ref{afinal},\ref{dfinal}) respectively, thus obtaining
\beqn
&& g_1^n(Q^2) = \int_0^1 dy~
\frac{y^{n-1}}{\left(1-\frac{y^2m^2}{Q^2}\right)^{n+1}}
\left[n^2 H_a(y)+\frac{2y^2 m^2}{Q^2}
~\left[nG_d(y)+n^2 H_d(y)\right]\right],
\nonumber \\
\label{g1n}
\\
&& g_2^n(Q^2) = n(n-1)\int_0^1 dy ~
\frac{y^{n-1}}{\left(1-\frac{y^2m^2}{Q^2}\right)^{n+1}}
\left[H_d(y)-H_a(y)-\frac{y^2m^2}{Q^2}H_d(y)\right].
\nonumber \\
\label{g2n}
\eeqn
Equations~(\ref{g1n},\ref{g2n}) are our main result. They express the moments
of the polarized structure functions $g_1$, $g_2$ in terms of matrix elements
of the operators appearing in the light-cone expansion of the forward 
scattering amplitude, at all orders in $m^2/Q^2$. Observe that when the 
twist-3 operator matrix elements $d_n$ are neglected, eqs.~(\ref{g1n},\ref{g2n})
obey the so-called Wandzura--Wilczek relation~\cite{WW}
\beq
g_2^n(Q^2)=-\frac{n-1}{n}g_1^n(Q^2) \;\;\;\;\;(for~d_n=0).
\label{wweq}
\eeq

The familiar $m=0$ 
result~\cite{kodaira} is easily recoverd by using 
eqs.~(\ref{fdef}-\ref{ghd}):
\beqn
&& g_1^n(Q^2) = a_n +{\cal O}\left(\frac{m^2}{Q^2}\right)
\label{g1n0}
\\
&& g_2^n(Q^2) = \frac{n-1}{n}(d_n-a_n)+{\cal O}\left(\frac{m^2}{Q^2}\right).
\label{g2n0}
\eeqn
The inverse Mellin transforms of eqs.~(\ref{g1n},\ref{g2n}) can be computed
as in ref.~\cite{GP}:
\beqn
g_1(x,Q^2)&=&\left(x^2\frac{d^2}{dx^2}+x\frac{d}{dx}\right)
\left[\frac{x}{r\xi}\left(H_a(\xi)+\frac{2\xi^2 m^2}{Q^2}H_d(\xi)\right)\right]
\nonumber\\
&&-\frac{2xm^2}{Q^2}\frac{d}{dx}\left[\frac{x\xi}{r}G_d(\xi)\right]
\label{g1x}
\\
g_2(x,Q^2)&=&x\frac{d^2}{dx^2}\left[\frac{x^2}{r\xi}
\left(\frac{\xi}{x}H_d(\xi)-H_a(\xi)\right)\right],
\label{g2x}
\eeqn
where 
\beq
\xi=\frac{2x}{1+r}; \;\;\;\; r=\sqrt{1+\frac{4x^2 m^2}{Q^2}}.
\eeq
An excellent test of the correctness of our calculation is a
comparison with the results of ref.~\cite{MU}.
We have checked that eliminating $g_1$ and $g_2$ from
eqs.~(18,19) of ref.~\cite{MU} using our eqs.~(\ref{g1x},\ref{g2x}) (a
non-trivial task) two identities are obtained. 

As in the unpolarized case, there is a well-known difficulty
in eqs.~(\ref{g1x},\ref{g2x}) at $x=1$: in fact, when $x=1$
the structure functions should vanish for kinematical reasons,
while the RHS of eqs.~(\ref{g1x},\ref{g2x}) are clearly nonzero, since
$\xi(x=1)<1$. This problem was discussed in ref.~\cite{DGP} for the unpolarized
case; the conclusion reached there is that in the large $x$ region dynamical
higher twist corrections become important and cannot be neglected any more.
This is because the twist $2+2k$ contribution to the $n^{th}$ moment
of a generic structure function has the form
\beq
B_{kn}(Q^2)\left(\frac{n\Lambda^2}{Q^2}\right)^k,
\label{ht}
\eeq
where $\Lambda$ is a mass scale of the order of a few hundreds MeV, and
the coefficients $B_{kn}(Q^2)$ have no power dependence on $n,k$ or $Q^2$.
The crucial feature of eq.~(\ref{ht}) is the presence of a factor $n^k$,
which arises because there are at least $n$ twist $2+2k$ operators
of a given dimension for each leading twist operator of the same dimension.
One can prove that the behaviour of structure functions in the
$x \sim 1$ region is governed by moments $n={\cal O}(Q^2/m^2)$;
in fact, when $x=1$ and $m^2/Q^2<<1$ we have
\beq
\xi\simeq 1-\frac{m^2}{Q^2};
\label{xi1}
\eeq
on the other hand, if we assume a $(1-\xi)^a$ behaviour for the structure
function, with 
$a$ of order 1, its $n^{th}$ moment receives the dominant contribution from
the region
\beq
\xi\simeq \frac{n}{a+n} \simeq 1-\frac{a}{n}.
\label{xi2}
\eeq
Comparing eqs.~(\ref{xi1}) and (\ref{xi2}), we obtain that the relevant 
moments for the $x=1$ region are of order
\beq
n=\frac{aQ^2}{m^2}
\eeq
as announced. Inserting this in eq.~(\ref{ht}), one immediately realizes that
the contribution of dynamical higher twists is no longer suppressed 
by inverse powers of $Q^2$ when $x$ is close to 1, and we cannot expect our
result, eqs.~(\ref{g1n},\ref{g2n}), to hold in this region.
A solution to this problem~\cite{Mir1,Mir2} is that of expanding the result
in powers of $m^2/Q^2$ up to any finite order. In this way, the dangerous
contribution of terms with large powers of $m^2/Q^2$ is not included.
The expansion remains reliable even when $Q^2$ is as low as $m^2$,
provided $x$ 
is not too large; in fact, powers of $m^2/Q^2$ always appear multiplied
by an equal power of $x^2$.
The expanded result of course cannot be
reliable at $x\simeq 1$, but this would not be the case even without expanding
in $m^2/Q^2$, since we are not including the contributions of eq.~(\ref{ht}),
which are important in this region.
Therefore, we will perform our phenomenological analysis expanding our results,
eqs.~(\ref{g1n},\ref{g2n}), to first order in $m^2/Q^2$. We have
\beqn
&&g_1^n(Q^2) = a_n+\frac{m^2}{Q^2}
\frac{n(n+1)}{(n+2)^2}\left(na_{n+2}+4d_{n+2}\right)
+{\cal O}\left(\frac{m^4}{Q^4}\right)
\label{g1n1}
\\ 
&&g_{2}^n(Q^2) = \frac{n-1}{n}\left(d_n-a_n\right)+\frac{m^2}{Q^2}
\frac{n(n-1)}{(n+2)^2}\left[nd_{n+2}-(n+1)a_{n+2}\right]
+ {\cal O}\left(\frac{m^4}{Q^4}\right).\nonumber\\
\label{g2n1}
\eeqn
We can now use eqs.~(\ref{g1n0}, \ref{g2n0}) to eliminate the matrix elements
$a_n,d_n$ from eqs.~(\ref{g1n1}, \ref{g2n1}) in favour of the moments of
$g_1$ and $g_2$ at zero nucleon mass, which we denote by
$g_{10}^n$ and $g_{20}^n$, respectively. We obtain
\beqn
&&g_1^n(Q^2) = g_{10}^n(Q^2)+\frac{m^{2}}{Q^{2}}\frac{n(n+1)}{(n+2)^{2}}
\left[(n+4)~g_{10}^{n+2}(Q^2)+4\frac{n+2}{n+1}~g_{20}^{n+2}(Q^2)\right]
+{\cal O}\left(\frac{m^4}{Q^4}\right)
\nonumber\\
\label{g1n1bis}
\\
&&g_2^n(Q^2) = g_{20}^n(Q^2)+\frac{m^{2}}{Q^{2}}\frac{n(n-1)}{(n+2)^{2}}
\left[n\frac{n+2}{n+1}g_{20}^{n+2}(Q^2)-g_{10}^{n+2}(Q^2)\right]
+ {\cal O}\left(\frac{m^4}{Q^4}\right).
\label{g2n1bis}
\eeqn

\section{Phenomenology}
Analyses of polarized deep-inelastic scattering data in the context of QCD at
next-to-leading order have been performed by different
groups~\cite{GehrmannStirling}-\cite{ABFR} in the zero-mass approximation. In
this section we will repeat the same analysis taking mass corrections into
account, in order to establish their practical importance. The quantities which
are directly measured in polarized deep-inelastic scattering experiments are the
asymmetries 
\beqn
&&A_\perp = \frac{\sigma^{\downarrow\rightarrow}
                 -\sigma^{\uparrow\rightarrow}}
                 {\sigma^{\downarrow\rightarrow}
                 +\sigma^{\uparrow\rightarrow}}
\label{aperp}
\\
&&A_\parallel = \frac{\sigma^{\uparrow\downarrow}
                     -\sigma^{\uparrow\uparrow}}
                     {\sigma^{\uparrow\downarrow}
                     +\sigma^{\uparrow\uparrow}},
\label{apar}
\eeqn
where the arrows refer to the orientation of the lepton and the proton spin
vectors with respect to the beam axis. One can show that
\beqn
&&A_\perp=d(A_2 -\zeta A_1)
\label{aperp2}
\\
&&A_\parallel=D(A_1 +\eta A_2),
\label{apar2}
\eeqn
where $A_1$ and $A_2$ are virtual photon cross section asymmetries,
and the coefficients $D,d,\eta,\zeta$ are fixed by the kinematics
of the process; in the target rest frame
\beq
D=\frac{E-\varepsilon E'}{E(1+\varepsilon R)},\;\;\;\;
\eta=\frac{\varepsilon\sqrt{Q^2}}{E-\varepsilon E'},\;\;\;\;
d = D\sqrt{\frac{2\varepsilon}{1+\varepsilon}}, \;\;\;\;
\zeta=\eta\frac{1+\varepsilon}{2\varepsilon}
\eeq
where $E$ ($E'$) is the energy of the incoming (outgoing) lepton, 
\beq
\frac{1}{\varepsilon}=1+2\left(1+\frac{Q^2}{4x^2m^2 }\right)
\tan^2\frac{\theta}{2}
\eeq
and $\theta$ is the lepton 
scattering angle. The ratio $R$ is the usual quantity
defined in unpolarized deep-inelastic scattering, namely
\beq
R=\frac{F_2}{2xF_1}\left(1+\frac{4x^2 m^2}{Q^2}\right)-1.
\label{R}
\eeq
The asymmetries $A_1$ and $A_2$ are directly related to polarized structure 
functions through
\beqn
&& A_1 = \frac{g_1 -\frac{4m^2 x^2}{Q^2}g_2}{F_1} 
\label{a1g1g2}
\\
&& A_2 = \frac{2mx}{\sqrt{Q^2}} \frac{g_1 +g_2}{F_1}. 
\label{a2g1g2}
\eeqn
Therefore, both $g_1$ and $g_2$ can be expressed in terms of the
measured asymmetries $A_\perp,A_\parallel$ and of unpolarized structure 
functions, through eqs.~(\ref{aperp2},\ref{apar2}) and
(\ref{a1g1g2},\ref{a2g1g2}).
When the nucleon mass is neglected, one has simply 
\beq
\label{zeromass}
\frac{A_\parallel}{D}=A_1=\frac{g_1}{F_1}=\frac{g_1}{F_2}2x(1+R)
\;\;\;\;(for~m=0)
\eeq
while in general this relationship involves both $A_\parallel$ and $A_\perp$.

The usual procedure for analysing data, adopted for example in
ref.~\cite{ABFR}, must be modified in different aspects. First, the term
proportional to $m^2/Q^2$ in eq.~(\ref{R}), which relates $F_1$ to $R$
and $F_2$, must now be included. Secondly, one must perform a global
fit of the measured asymmetries using
eq.~(\ref{g1n1bis}) (if the analysis is done in moment space),
where the moments $g^n_{10}$ are given in terms
of moments of the coefficient functions
and of the polarized parton distributions, which by definition
are proportional to the matrix elements $a_n$.
A difficulty immediatly arises, because
moments of the structure function $g_{20}$ also appear in eq.~(\ref{g1n1bis});
therefore, it is not possible to treat $g_1$ and $g_2$ independently,
as in the $m=0$ case.
One could in principle circumvent this problem using experimental information
on $g_2$; unfortunately, $g_2$ data available up to now are restricted
to a very limited range of $x$ and $Q^2$, and are affected by large
uncertainties~\cite{g2data}.
We prefer here to follow a different strategy. We will perform fits
to data
in two different ways, characterized by two different assumptions on $g_{20}$:
either we simply set $g_{20}=0$, or we relate $g_{20}$ to $g_{10}$ by the
Wandzura-Wilczek relation,
\beq
g_{20}^{n+2}(Q^2)=-\frac{n+1}{n+2}g_{10}^{n+2}(Q^2).
\label{wweq0}
\eeq
Notice that none of the two 
assumptions is theoretically justified: there is of course no reason to assume 
that $g_{20}$ vanishes, nor that twist-3 operators give a negligible 
contribution. However, both assumptions are consistent with presently available
information on $g_2$. We will check that this procedure actually allows
to make a reliable estimate of the effect of mass corrections.
Notice that some of the experimental
collaborations present values of the asymmetry $A_1$,
while others give values of the combination of $A_\parallel$ and
$A_\perp$ which corresponds to $g_1/F_1$. The two quantities coincide for
$m=0$, as already observed, but they do not when mass corrections are
included; so in the first case the asymmetry must be fitted by
$[g_1-(4m^2x^2/Q^2)g_2]/F_1$, in the second case simply by $g_1/F_1$.

The structure function $g_{10}$ is related to the polarized quark and gluon
distributions by
\beq
g_{10}(x,Q^2)=\frac{\langle e^2 \rangle}{2} \left[C_{NS}\otimes \Delta q_{NS}
+C_S\otimes \Delta \Sigma + 2n_fC_g \otimes \Delta g\right],
\label{1}
\eeq
where $\langle e^2 \rangle=n_f^{-1}\sum_{i=1}^{n_f} e^2_i$, $\otimes$ denotes
convolution with respect to $x$, and the nonsinglet and singlet quark
distributions are defined as
\beq
\Delta q_{NS}\equiv \sum_{i=1}^{n_f}
\left(\frac{e^2_i}{\langle e^2 \rangle} - 1\right)
(\Delta q_i+\Delta \bar q_i),\qquad
\Delta \Sigma\equiv \sum_{i=1}^{n_f}(\Delta q_i+\Delta \bar q_i),
\label{2}
\eeq
where $\Delta q_i$ and $\Delta \bar q_i$ are the quark and antiquark
distributions
of flavor $i$ and $\Delta g$ is the polarized gluon distribution. 

A first fit (called fit A in ref.~\cite{ABFR}) is performed
with the initial parton distributions parametrized
at $Q_0^2 = 1$~GeV$^2$ according to the conventional form
\beq
\Delta f(x,Q_0^2) = {\cal N}_f \eta_f x^{\alpha_f} (1-x)^{\beta_f}
(1 + \gamma_fx^{\delta_f})
\label{firstclass}
\eeq
where $\Delta f$ denotes $\Delta q_{NS}$, $\Delta \Sigma$ or
$\Delta g$; the factor ${\cal N}_f$ is chosen so that the first moment of
$\Delta f$ is equal to $\eta_f$. We have fixed
\beq
\delta_{\Sigma}=\delta_g=1,\qquad\delta_{NS}=0.75,\qquad\beta_g=4,
\qquad\gamma_{\Sigma}=\gamma_g\qquad {\rm (fit~A).}
\label{fitapar}
\eeq
In a second class of fits (fit B) the rise at small $x$ is at most
logarithmic:
\beqn
\Delta \Sigma &=& {\cal N}_{\Sigma}
\eta_{\Sigma}x^{\alpha_{\Sigma}}
\left(\log 1/x\right)^{\beta_{\Sigma}}\nonumber\\
\Delta q_{NS} &=& {\cal N}_{NS} \eta_{NS}
\left[\left(\log 1/x\right)^{\alpha_{NS}} +\gamma_{NS}x
\left(\log1/x\right)^{\beta_{NS}}\right] \qquad {\rm (fit~B),}\\
\Delta g &=& {\cal N}_g \eta_g \left[\left(\log 1/x\right)^{\alpha_g}
+\gamma_g x\left(\log 1/x\right)^{\beta_g}\right]
\nonumber \label{fitb}
\eeqn
The motivations for the choice of these particular parametrizations are 
discussed in ref.~\cite{ABFR}.

\begin{figure}
\centerline{
   \epsfig{figure=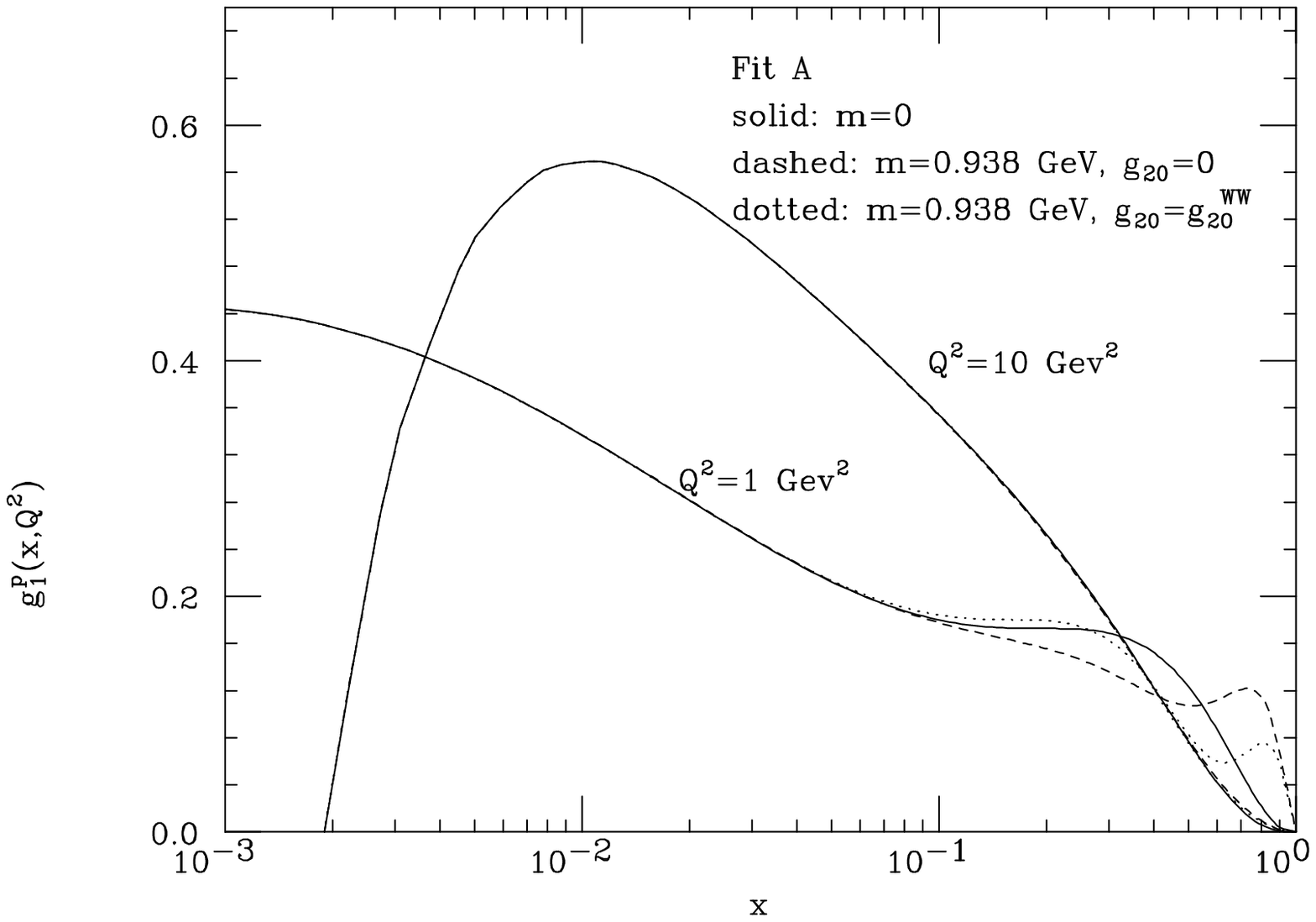,width=0.48\textwidth,clip=}
   \hfill
   \epsfig{figure=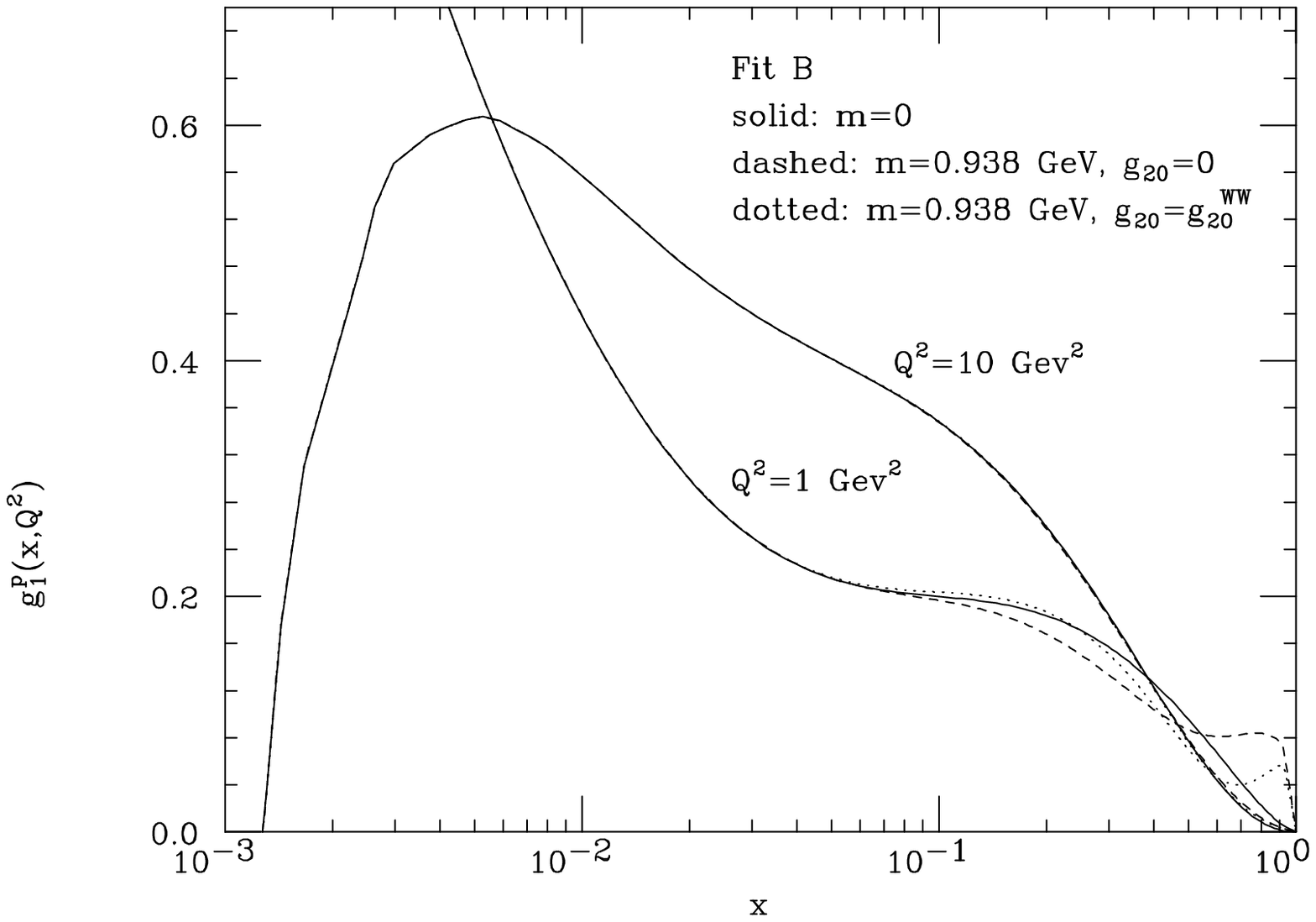,width=0.48\textwidth,clip=} }
\ccaption{}{ \label{g1p} The structure function $g_1$ for the proton
at $Q^2=1,10$~GeV$^2$ in fits A and B, with and without target mass
corrections (parameters fixed at the $m=0$ values).
}
\end{figure}

We have first performed fits~A and~B with fixed $\as(m_Z)=0.118$ for $m=0$,
and with the obtained values of the fit parameters we have recomputed
the structure function $g_1$ using the mass-corrected formula,
eq.~(\ref{g1n1bis}).
The effect of target mass corrections turns out to be
very small. As an example, in fig.~\ref{g1p} we show 
$g_1(x,Q^2)$ for the proton at $Q^2=1$~GeV$^2$
and $Q^2=10$~GeV$^2$ and $m=0$ for both fits A and B; in the same plot,
we also show
$g_1$ obtained including mass corrections, using the same values of the fit
parameters found for $m=0$, for both assumptions $g_{20}=0$ and 
$g_{20}=g_{20}^{WW}$.
Differences among the three
curves for $Q^2=1$~GeV$^2$ are sizeable only in the large-$x$ region,
while at $Q^2=10$~GeV$^2$ the three curves are practically undistinguishable
on this scale.
%%%%%%%%%%%%%%%%%%%%%%%%%%%%%%%%%%%%%%%%%%%%%%%%%%%%%%%%%%%%%%%%%%%%%%
%%%%%%%%%%%%%%%%%%%%%%%%%%%%%%%%%%%%%%%%%%%%%%%%%%%%%%%%%%%%%%%%%%%%%%

\begin{figure}
\centerline{
   \epsfig{figure=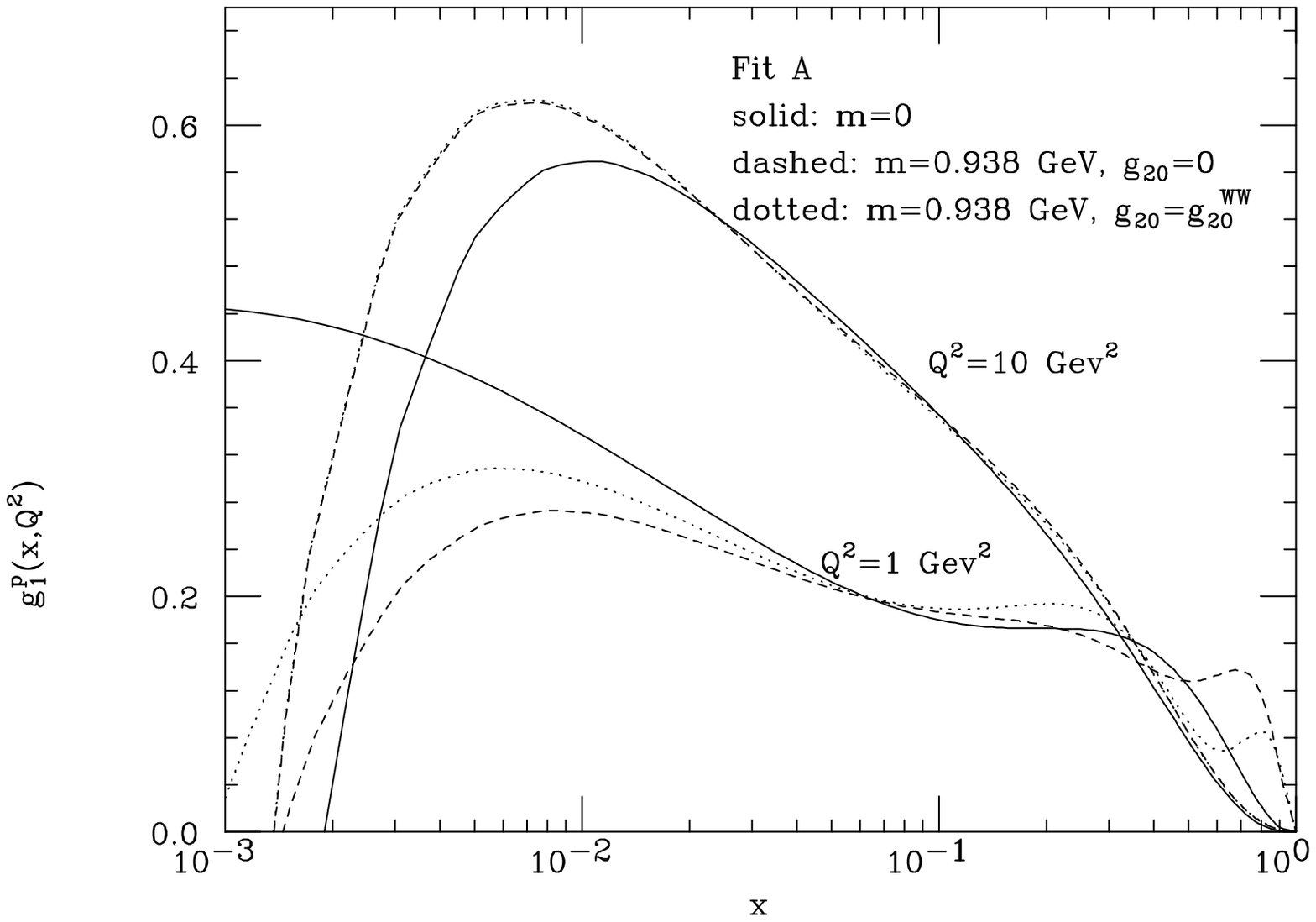,width=0.48\textwidth,clip=}
   \hfill
   \epsfig{figure=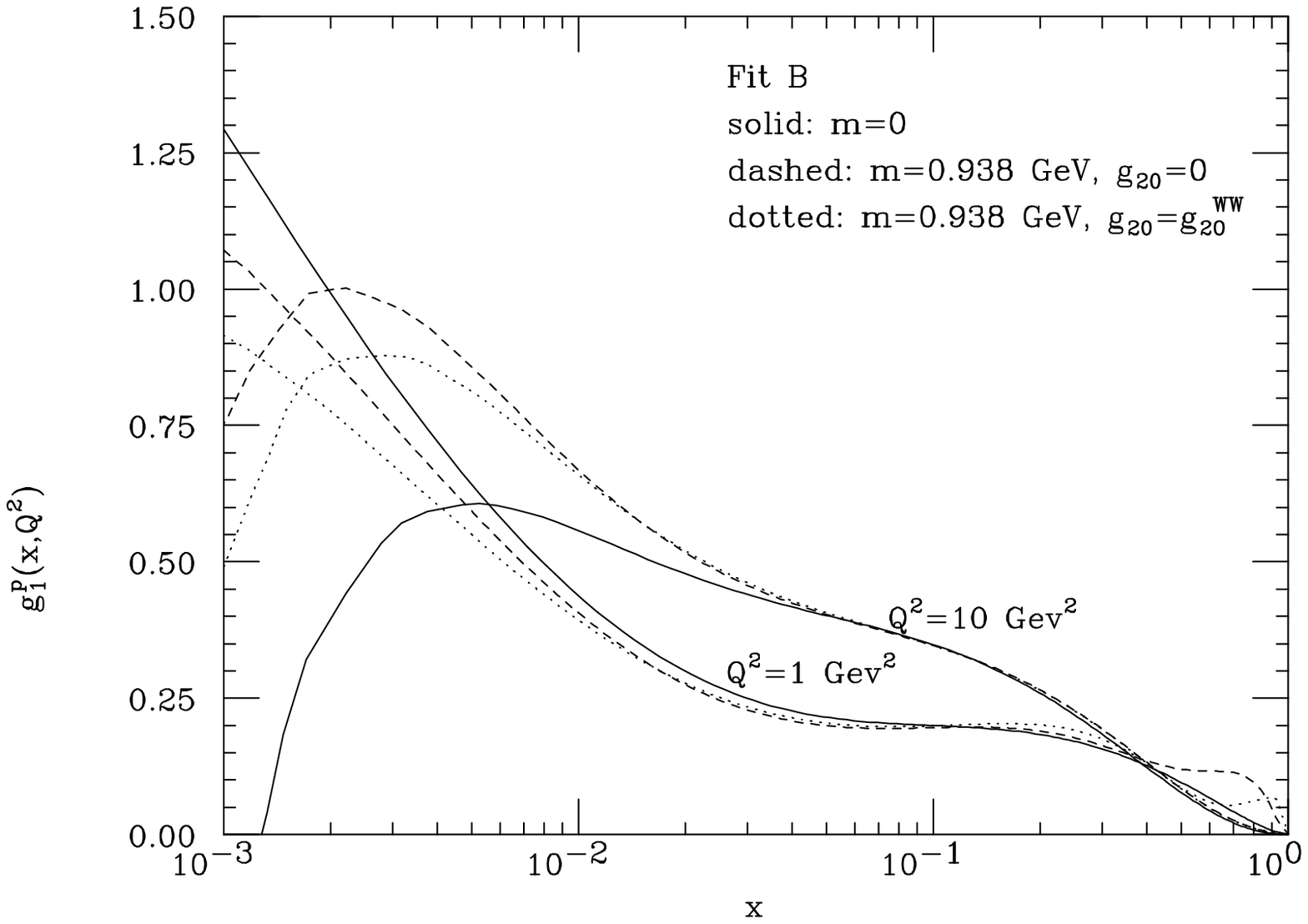,width=0.48\textwidth,clip=} }
\ccaption{}{ \label{g1p_fit} The structure function $g_1$ for the proton
at $Q^2=1,10$~GeV$^2$ in fits A and B, with and without target mass
corrections (fitted parameters).
}
\end{figure}

We have then repeated the same fits for $m=0.938$~GeV
with both assumptions on $g_{20}$, $g_{20}=0$ and $g_{20}=g_{20}^{WW}$.
The results for $g_1$ (proton) are shown in fig.~\ref{g1p_fit}.
%%%%%%%%%%%%%%%%%%%%%%%%%%%%%%%%%%%%%%%%%%%%%%%%%%%%%%%%%%%%%%%%%%%%%%
%%%%%%%%%%%%%%%%%%%%%%%%%%%%%%%%%%%%%%%%%%%%%%%%%%%%%%%%%%%%%%%%%%%%%%
Observe that also in this case the difference between the solid curves,
which correspond to the $m=0$ case, and the dashed and dotted curves,
corresponding to $m\neq 0$, are quite large at $Q^2=1$~GeV$^2$ for
$x> 0.2$; when $x$ is very large ($x>0.7$), however, the
effect is not physical because in this region
the approximation under which mass corections have been computed
is not reliable. However, these differences
become negligible at higher vales of $Q^2$ because of the $m^2/Q^2$
suppression factor, as one can see by looking at the $Q^2=10$~GeV$^2$
curves; this guarantees that the uncertainty of our calculation
connected with the expansion in $m^2/Q^2$ at first order
has little effect on the results of our fits to data, since data at large $x$
are taken at high values of $Q^2$.
The effect at small $x$, $x<0.01$, is also rather large;
however in this region there are
few or no data and thus $g_{10}$
itself is affected by a substantial uncertainty.

The results of fits A and B for $m\neq 0$\footnote{The results in the
first column of table~\ref{fitga}
are slightly different from those of ref.~\cite{ABFR} for two reasons:
first, we are fixing $\beta_g=4$ in fit A, second, we are using
more recent data sets for proton SMC data and neutron E154 data.}
for some of the fitted parameters
are shown in the second and third columns of table~\ref{fitga}.
The errors quoted in the table
are statistical errors originating from experimental uncertainties only. 
%%%%%%%%%%%%%%%%%%%%%%%%%%%%%%%%%%%%%%%%%%%%%%%%%%%%%%%%%%%%%%%%%%%%%%%%
\begin{center}
\begin{table}
\vspace*{0.5cm}
\begin{center}
\begin{tabular} {|l|c|c|c|}
\hline 
\multicolumn{4}{|c|} {FIT A} \\
\hline 
                  & $m=0$            & $g_{20}=0$       & $g_{20}=g_{20}^{WW}$\\
\hline
\hline
$g_A$             & $1.167\pm 0.045$ & $1.192\pm 0.040$ & $1.200\pm 0.043$\\
$\eta_\Sigma$     & $0.426\pm 0.037$ & $0.408\pm 0.027$ & $0.416\pm 0.031$\\
$\eta_g$          & $0.98 \pm 0.25$  & $0.81 \pm 0.35$  & $0.83 \pm 0.32$ \\
$a_0$(10 GeV$^2$) & $0.19 \pm 0.04$  & $0.21 \pm 0.08$  & $0.21 \pm 0.07$ \\
\hline
d.o.f.            & 114-10           & 114-10           & 114-10          \\
$\chi^2$          & 91.2             & 86.9             & 89.5            \\
$\chi^2$/d.o.f.   & 0.88             & 0.84             & 0.86            \\
\hline
\end{tabular}
\vspace*{0.5cm}
\begin{tabular} {|l|c|c|c|}
\hline 
\multicolumn{4}{|c|} {FIT B} \\
\hline 
                  & $m=0$            & $g_{20}=0$       & $g_{20}=g_{20}^{WW}$\\
\hline
\hline
$g_A$             & $1.253\pm 0.057$ & $1.292\pm 0.056$ & $1.277\pm 0.058$ \\
$\eta_\Sigma$     & $0.455\pm 0.038$ & $0.423\pm 0.034$ & $0.428\pm 0.037$ \\
$\eta_g$          & $1.40 \pm 0.32$  & $1.00 \pm 0.33$  & $0.99 \pm 0.35$  \\
$a_0$(10 GeV$^2$) & $0.13 \pm 0.05$  & $0.18 \pm 0.06$  & $0.19 \pm 0.06$  \\
\hline
d.o.f.            & 114-11           & 114-11           & 114-11           \\
$\chi^2$          & 86.8             & 86.3             & 90.1             \\
$\chi^2$/d.o.f.   & 0.84             & 0.84             & 0.87             \\
\hline
\end{tabular}
\end{center}
\ccaption{}{\label{fitga}Results of fits A and B with fixed $\as(m_Z)=0.118$.}
\end{table}
\end{center}
%%%%%%%%%%%%%%%%%%%%%%%%%%%%%%%%%%%%%%%%%%%%%%%%%%%%%%%%%%%%%%%%%%%%%%%%
In ref.~\cite{ABFR} the uncertainty on $g_A$ from higher twist corrections
was estimated to be $\pm 0.03$. We can see from table~\ref{fitga}
that target mass corrections actually induce deviations of this order from
the $m=0$ value for $g_A$. Deviations are always in the direction
of making $g_A$ larger than in the $m=0$ case. We observe also that the values
of $g_A$ obtained with the two different assumptions for 
$g_{20}$ are quite close
to each other, thus suggesting that the assumed form of $g_{20}$ has a small
impact on the final results.

We have also performed fits A and B using $\as$ as a free parameter, with $g_A$
fixed to its measured value $g_A=1.2573$. The results are shown in
table~\ref{fitas}. Also in this case, mass corrections induce variations on
$\as(m_Z)$ which are compatible with the higher twist uncertainty of $\pm 0.004$
estimated in ref.~\cite{ABFR}. 
%%%%%%%%%%%%%%%%%%%%%%%%%%%%%%%%%%%%%%%%%%%%%%%%%%%%%%%%%%%%%%%%%%%%%%%%
\begin{center}
\begin{table}
\vspace*{0.5cm}
\begin{center}
\begin{tabular} {|l|c|c|c|}
\hline 
\multicolumn{4}{|c|} {FIT A} \\
\hline 
                  & $m=0$            & $g_{20}=0$       & $g_{20}=g_{20}^{WW}$\\
\hline
\hline
$\as(m_Z)$        & $0.118\pm 0.005$ & $0.117\pm 0.004$ & $0.120\pm 0.003$\\
$\eta_\Sigma$     & $0.433\pm 0.039$ & $0.415\pm 0.027$ & $0.423\pm 0.028$\\
$\eta_g$          & $1.04 \pm 0.45$  & $0.91 \pm 0.29$  & $0.85 \pm 0.40$ \\
$a_0$(10 GeV$^2$) & $0.18 \pm 0.08$  & $0.20 \pm 0.09$  & $0.20 \pm 0.09$ \\
\hline
d.o.f.            & 114-10           & 114-10           & 114-10          \\
$\chi^2$          & 94.4             & 89.1             & 90.9            \\
$\chi^2$/d.o.f.   & 0.91             & 0.86             & 0.87 \\
\hline
\end{tabular}
\vspace*{0.5cm}
\begin{tabular} {|l|c|c|c|}
\hline 
\multicolumn{4}{|c|} {FIT B} \\
\hline 
                  & $m=0$            & $g_{20}=0$       & $g_{20}=g_{20}^{WW}$\\
\hline
\hline
$\as(m_Z)$        & $0.123\pm 0.003$ & $0.118\pm 0.005$ & $0.121\pm 0.004$ \\
$\eta_\Sigma$     & $0.448\pm 0.036$ & $0.407\pm 0.036$ & $0.418\pm 0.033$ \\
$\eta_g$          & $1.01 \pm 0.32$  & $0.73 \pm 0.33$  & $0.72 \pm 0.31$  \\
$a_0$(10 GeV$^2$) & $0.14 \pm 0.07$  & $0.22 \pm 0.07$  & $0.21 \pm 0.06$  \\
\hline
d.o.f.            & 114-11           & 114-11           & 114-11           \\
$\chi^2$          & 84.8             & 86.3             & 89.2             \\
$\chi^2$/d.o.f.   & 0.82             & 0.84             & 0.87             \\
\hline
\end{tabular}
\end{center}
\ccaption{}{\label{fitas}Results of fits A and B with fixed $g_A=1.2573$.}
\end{table}
\end{center}
%%%%%%%%%%%%%%%%%%%%%%%%%%%%%%%%%%%%%%%%%%%%%%%%%%%%%%%%%%%%%%%%%%%%%%

The values of the first moment of the quark singlet distribution, $\eta_\Sigma$,
and of the first moment of the gluon distribution at $Q^2=1$~GeV$^2$, $\eta_g$,
are also shown in tables~\ref{fitga}-\ref{fitas}. Also for these quantities, the
assumption on $g_{20}$ has little effect. The introduction of mass terms tend to
give smaller values for $\eta_\Sigma$ and $\eta_g$, but deviations from the
values of the $m=0$ case are all within statistical errors. 

In tables~\ref{fitga}-\ref{fitas} we also present values of the singlet axial
charge 
\beq
a_0(Q^2) = \int_0^1dx\,\left[\Delta \Sigma(x,Q^2) - n_f 
\frac{\as(Q^2)}{2\pi}\Delta g(x,Q^2)\right]
\label{azero}
\eeq
for $Q^2=10$~GeV$^2$. Values of $a_0$ for $m\not=0$ are slightly larger than in
the massless fits. The only case in which a sizeable devitation from the $m=0$
result is observed is fit B with $g_A$ fixed. Also in this case, however, the
difference is compatible with the statistical error. 

\begin{figure}
\centerline{
   \epsfig{figure=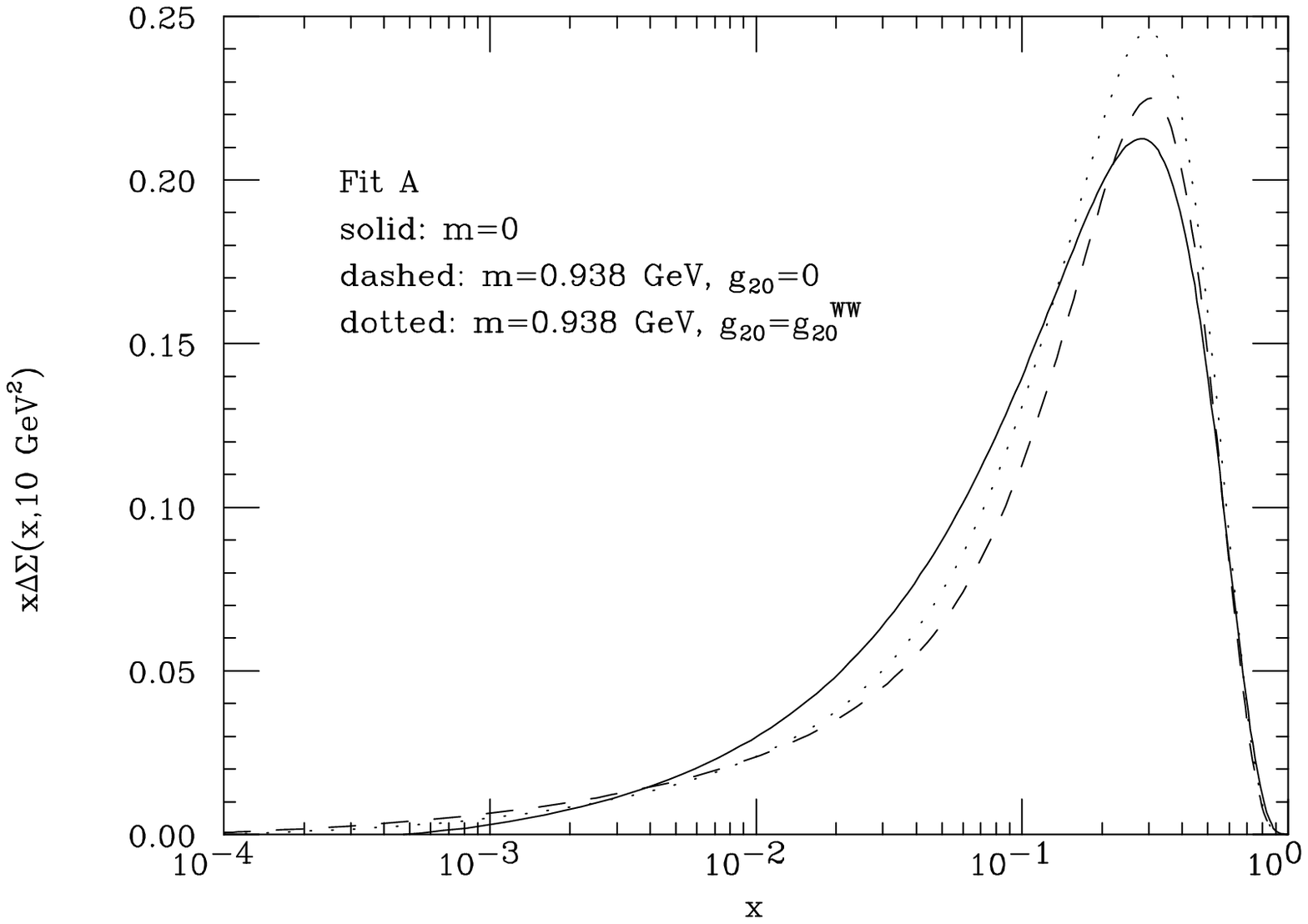,width=0.48\textwidth,clip=}
   \hfill
   \epsfig{figure=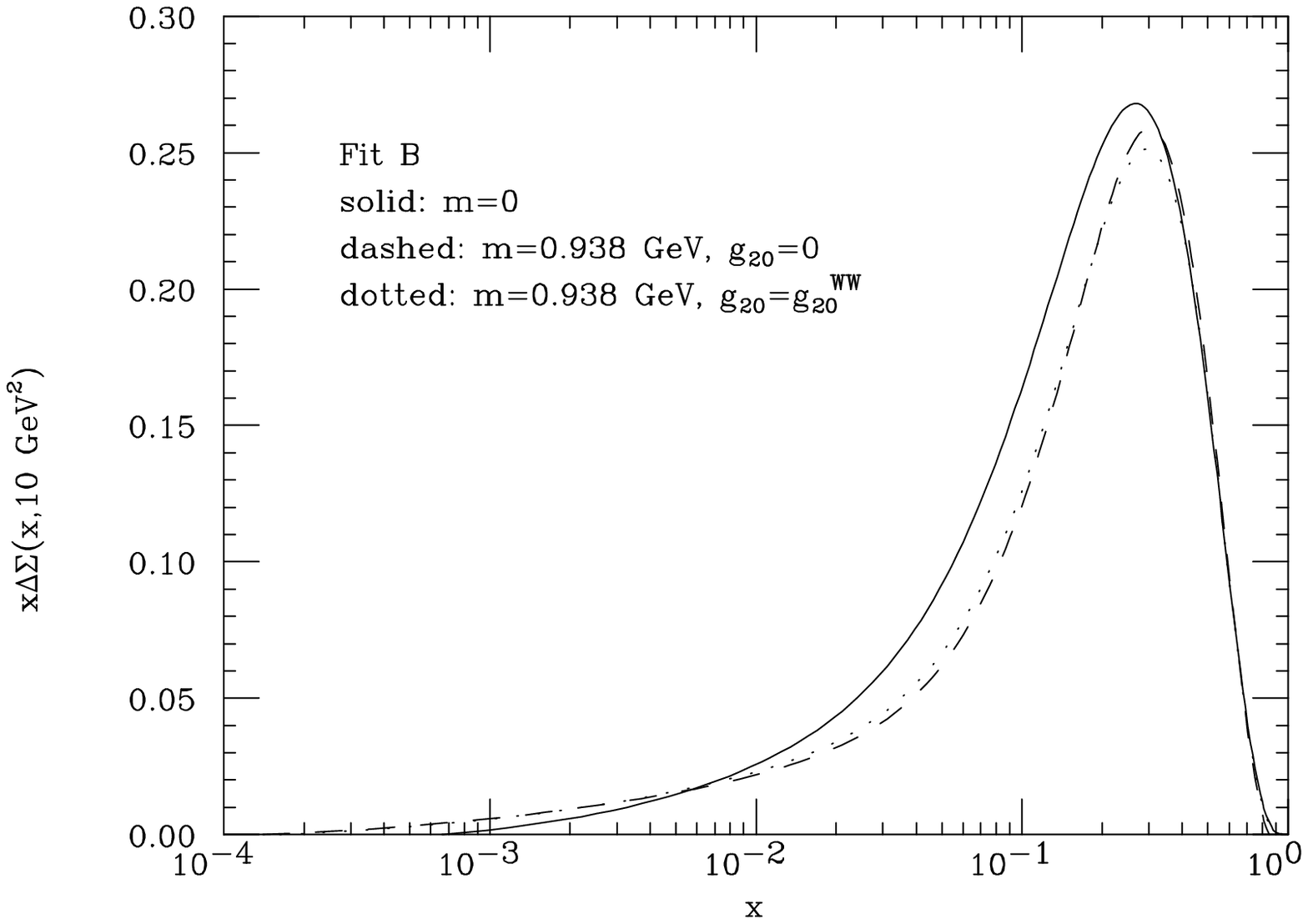,width=0.48\textwidth,clip=} }
\ccaption{}{ \label{singlet} The quark singlet distribution at
$Q^2=10$~GeV$^2$ in fits A and B, with and without target mass corrections.
}
\end{figure}

\begin{figure}
\centerline{
   \epsfig{figure=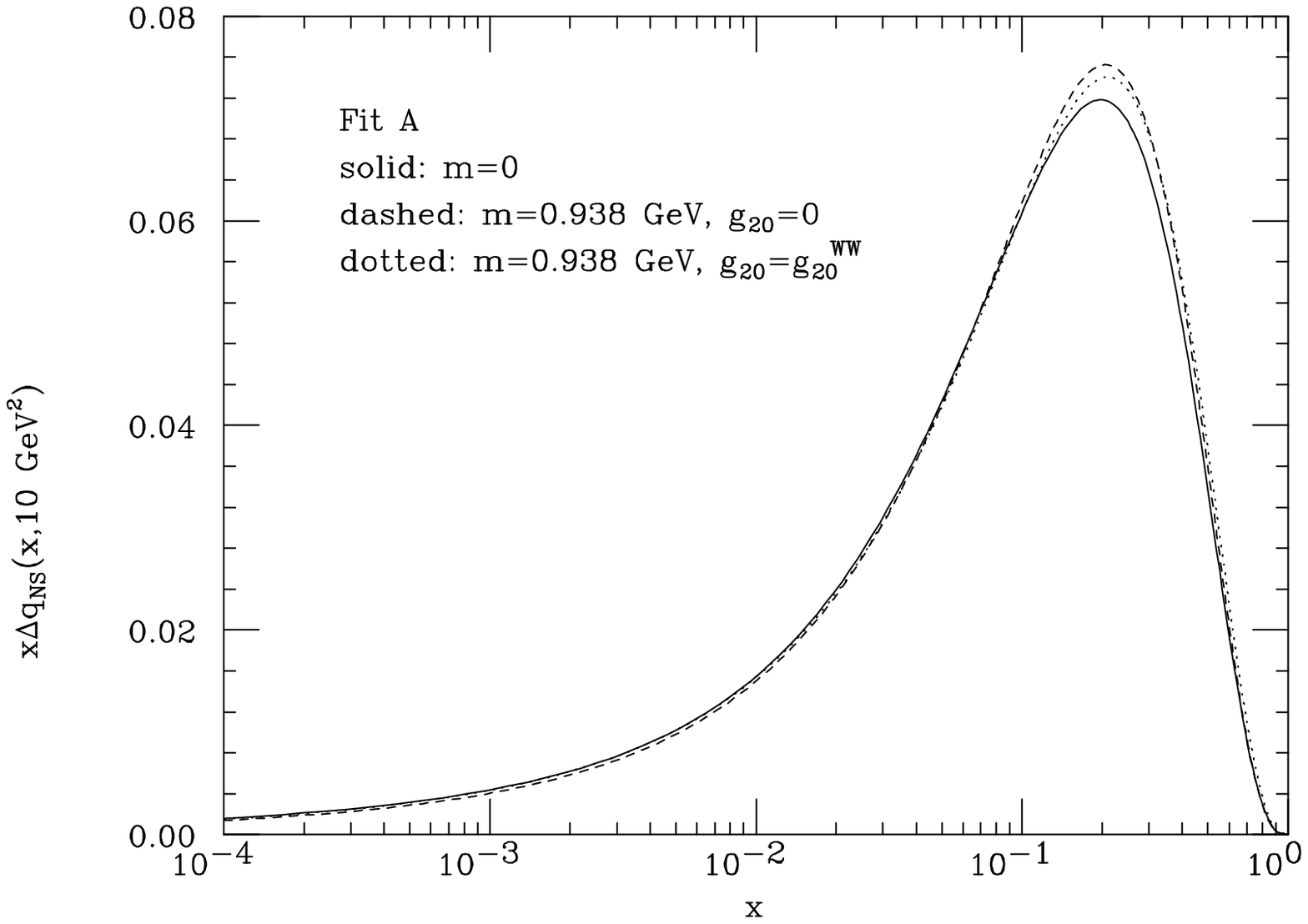,width=0.48\textwidth,clip=}
   \hfill
   \epsfig{figure=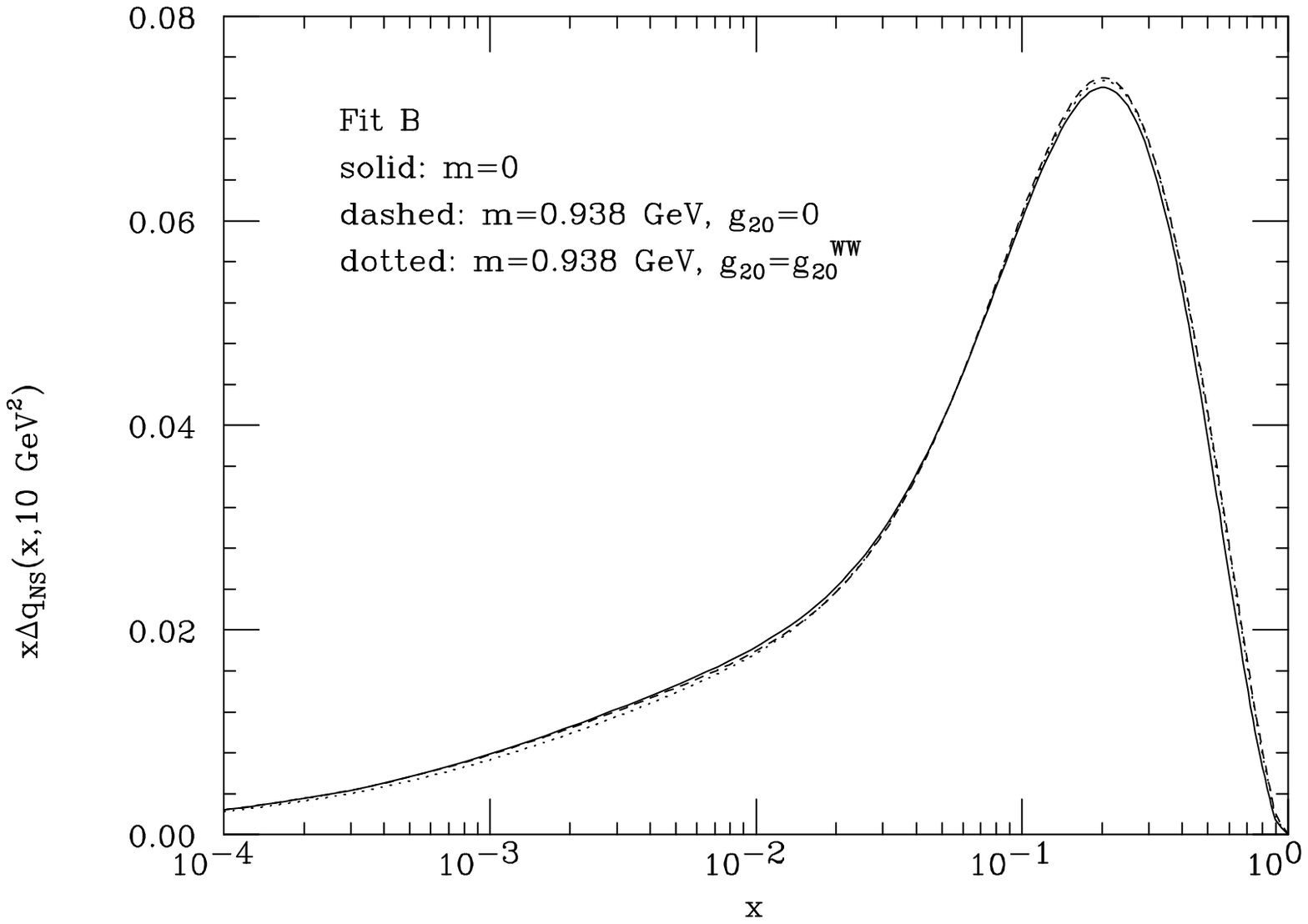,width=0.48\textwidth,clip=} }
\ccaption{}{ \label{nonsinglet} The quark nonsinglet distribution for the
proton at
$Q^2=10$~GeV$^2$ in fits A and B, with and without target mass corrections.
}
\end{figure}

\begin{figure}
\centerline{
   \epsfig{figure=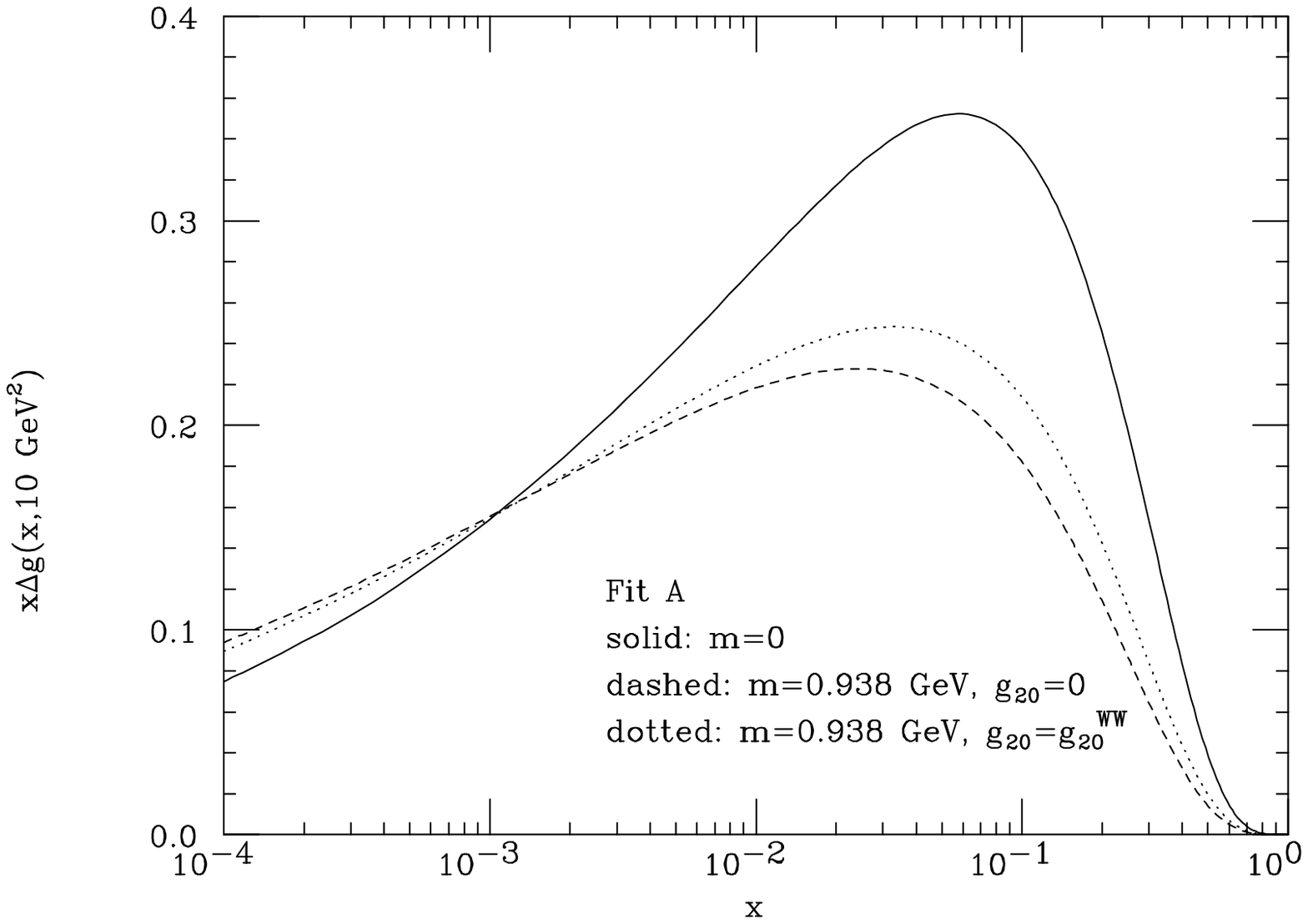,width=0.48\textwidth,clip=}
   \hfill
   \epsfig{figure=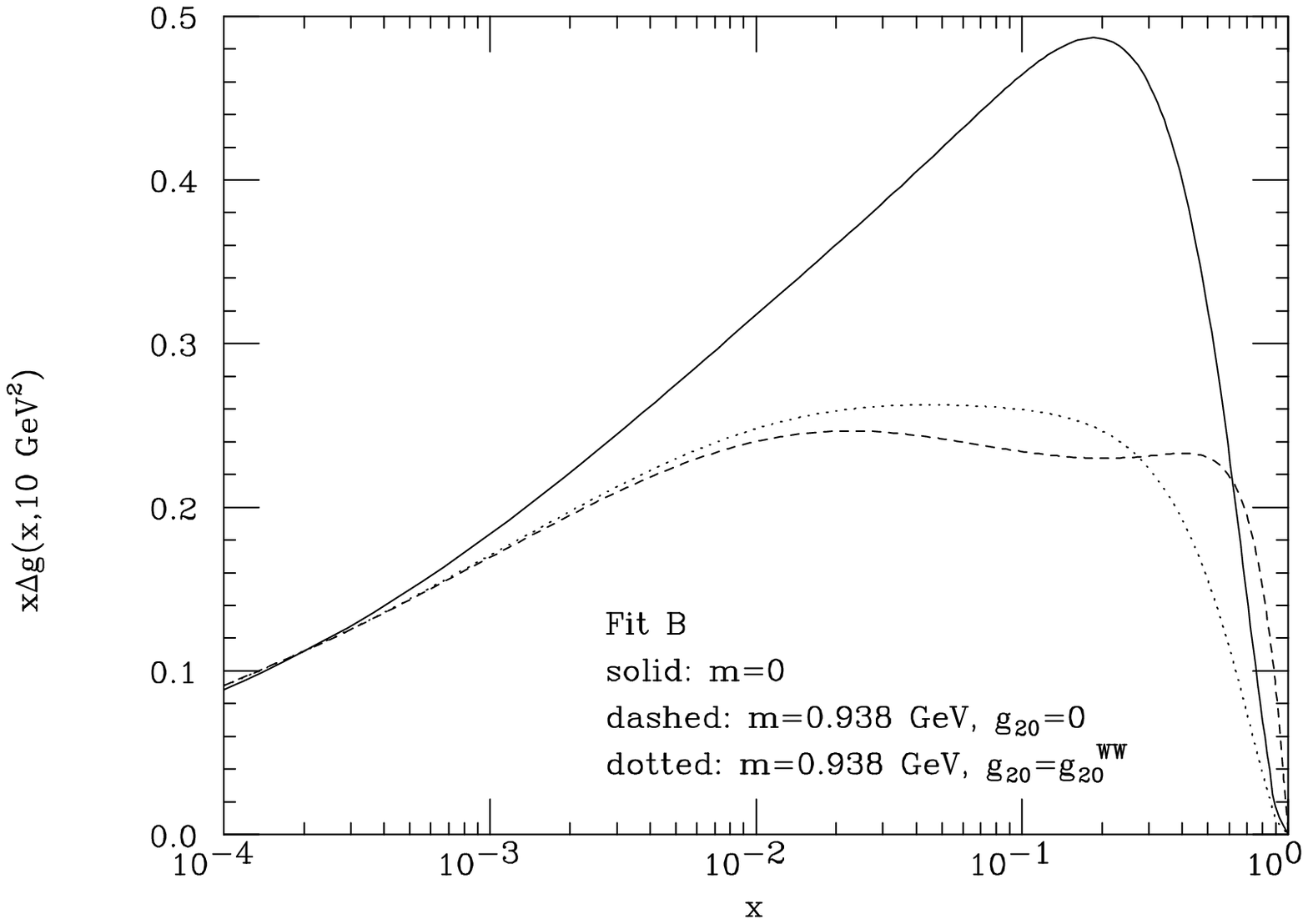,width=0.48\textwidth,clip=} }
\ccaption{}{ \label{gluon}The gluon distribution at
$Q^2=10$~GeV$^2$ in fits A and B, with and without target mass corrections.
}
\end{figure}

Finally, we compare the polarized parton distribution functions for quark
singlet, quark non-singlet and gluon obtained with and without mass corrections.
They are shown, for $Q^2=10$~GeV$^2$ in figs.~\ref{singlet}, \ref{nonsinglet}
and \ref{gluon} for fits A and B with fixed $\as(m_Z)=0.118$. 
%%%%%%%%%%%%%%%%%%%%%%%%%%%%%%%%%%%%%%%%%%%%%%%%%%%%%%%%%%%%%%%%%%%%%%
%%%%%%%%%%%%%%%%%%%%%%%%%%%%%%%%%%%%%%%%%%%%%%%%%%%%%%%%%%%%%%%%%%%%%% 
%%%%%%%%%%%%%%%%%%%%%%%%%%%%%%%%%%%%%%%%%%%%%%%%%%%%%%%%%%%%%%%%%%%%%%
%%%%%%%%%%%%%%%%%%%%%%%%%%%%%%%%%%%%%%%%%%%%%%%%%%%%%%%%%%%%%%%%%%%%%%
As expected from the above discussion, curves for 
$g_{20}=0$ and $g_{20}=g_{20}^{WW}$ are
quite close to each other. Quark distributions are
hardly affected by mass corrections.
The polarized gluon density $\Delta g$ is determined, in this procedure, only
through the effect of scaling violations. It is therefore
less constrained than quark distributions.
However, the modifications of the gluon
distribution induced by mass corrections,
although sizeable, are considerably less important than the uncertainty
originated by the choice of different functional forms
for the input distribution (see ref.~\cite{ABFR}).
%%%%%%%%%%%%%%%%%%%%%%%%%%%%%%%%%%%%%%%%%%%%%%%%%%%%%%%%%%%%%%%%%%%%%%
%%%%%%%%%%%%%%%%%%%%%%%%%%%%%%%%%%%%%%%%%%%%%%%%%%%%%%%%%%%%%%%%%%%%%%

\section{Conclusions}
We have computed 
target mass corrections to nucleon structure functions
in polarized deep-inelastic scattering. Our results are consistent with
those obtained in ref.~\cite{MU} using a different technique.

Target mass corrections
can in principle be important in the context of polarized
deep-inelastic scattering, because part of the data are
taken at relatively low values of $Q^2$. For this reason, we 
have performed an analysis of all available polarized
deep-inelastic scattering data in the framework of
perturbative QCD, including the contribution
of target mass corrections up to terms of order $m^2/Q^2$. We have
proved that this approximation is reliable in the kinematical range of
presently available data. We have found that the
effect of mass corrections is generally small;
for example the value of the axial coupling $g_A$ 
is enlarged by approximately $0.03$ when target mass corrections
are included.
The strong coupling constant $\as(m_Z)$ receives corrections
of approximately $0.004$. Both deviations are
compatible with higher twist uncertainties estimated in previous works.

Quark distributions are practically unchanged by the introduction of mass terms.
The polarized gluon distributions, which is only determined through scaling
violations, is comparatively more affected by mass corrections,
but also in this case the values of the fitted parameters
do not deviate from those obtained for $m=0$ by more than one
standard deviation.

\section*{Acknowledgements}
We are greatly indebted to Stefano Forte for many useful discussions and
suggestions.
We also thank G.~Altarelli, R.~Ball and S.~Frixione
for carefully reading the manuscript, and T.~Uematsu
for discussions.

One of us (GR) thanks the RIKEN-BNL Research Center for the kind
hospitality while completing this work.

\vfill\eject

\end{document}